\documentclass{article}

\usepackage[preprint]{neurips_2026}
\usepackage{colortbl}
\usepackage{tabularx}
\usepackage{tcolorbox}
\tcbuselibrary{skins}
\usepackage{array}
\usepackage{ragged2e}

\usepackage[utf8]{inputenc} % allow utf-8 input
\usepackage[T1]{fontenc}    % use 8-bit T1 fonts
\usepackage{hyperref}       % hyperlinks
\usepackage{url}            % simple URL typesetting
\usepackage{booktabs}       % professional-quality tables
\usepackage{amsfonts}       % blackboard math symbols
\usepackage{nicefrac}       % compact symbols for 1/2, etc.
\usepackage{microtype}      % microtypography
\usepackage{xcolor}         % colors
\usepackage{amsmath}
\usepackage{multirow}
\usepackage[capitalize,noabbrev]{cleveref}
\usepackage[textsize=tiny]{todonotes}
\newcommand{\legendswatch}[1]{\textcolor{#1}{\rule{0.8em}{0.8em}}}

\title{Do not copy and paste!\\
Rewriting strategies for code retrieval.}

\author{%
  Andrea~Gurioli \\
  DISI\\
  University of Bologna\\
  \texttt{andrea.gurioli5@unibo.it} \\
 \And
  Federico~Pennino \\
  DISI\\
  University of Bologna\\
  \texttt{federico.pennino2@unibo.it} \\
\And
  Maurizio~Gabbrielli \\
  DISI\\
  University of Bologna\\
  \texttt{maurizio.gabbrielli@unibo.it} \\
}

\begin{document}

\maketitle

\begin{abstract}
 Embedding-based code retrieval often suffers when encoders overfit
  to surface syntax. Prior work mitigates this by using LLMs to
  rephrase queries and corpora into a normalized style, but leaves
  two questions open: \emph{how much} representational shift helps,
  and \emph{when} is the per-query LLM call justified? We study a
  hierarchy of three rewriting strategies---stylistic rephrasing,
  NL-enriched PseudoCode, and full Natural-Language transcription---%
  under joint query--corpus (QC, online) and corpus-only (C, offline)
  augmentation, across six CoIR benchmarks, five encoders, and three
  rewriters spanning independent model families (Qwen, DeepSeek,
  Mistral). We are the first to evaluate NL-enriched PseudoCode and
  snippet-level Natural Language as \emph{direct} retrieval
  representations, rather than as transient intermediates. Full NL
  rewriting with QC yields the largest gains ($+0.51$ absolute
  NDCG@10 on CT-Contest for MoSE-18), while corpus-only rewriting
  degrades retrieval in $56$ of $90$ configurations (${\sim}62\%$).
  We introduce two diagnostics, $\Delta H$ (token entropy) and
  $\Delta\bar{s}$ (embedding cosine), and show that $\Delta H$
  predicts retrieval gain under QC across all three rewriter
  families (pooled Spearman $\rho{=}{+}0.436$, $p{<}0.001$ on
  DeepSeek+Codestral; $\rho{=}{+}0.593$ on Codestral alone;
  $\rho{=}{+}0.356$ on Qwen). This establishes $\Delta H$ as a
  cheap, rewriter-agnostic proxy for deciding when rewriting pays
  off \emph{before} running retrieval. Our analysis reframes LLM
  rewriting as a cost--benefit decision: it is most effective as a
  remediation layer for lightweight encoders on code-dominant
  queries, with diminishing returns for strong encoders or NL-heavy
  queries.
\end{abstract}

\section{Introduction}
\begin{figure}[!ht]
  \centering
  \includegraphics[width=\linewidth]{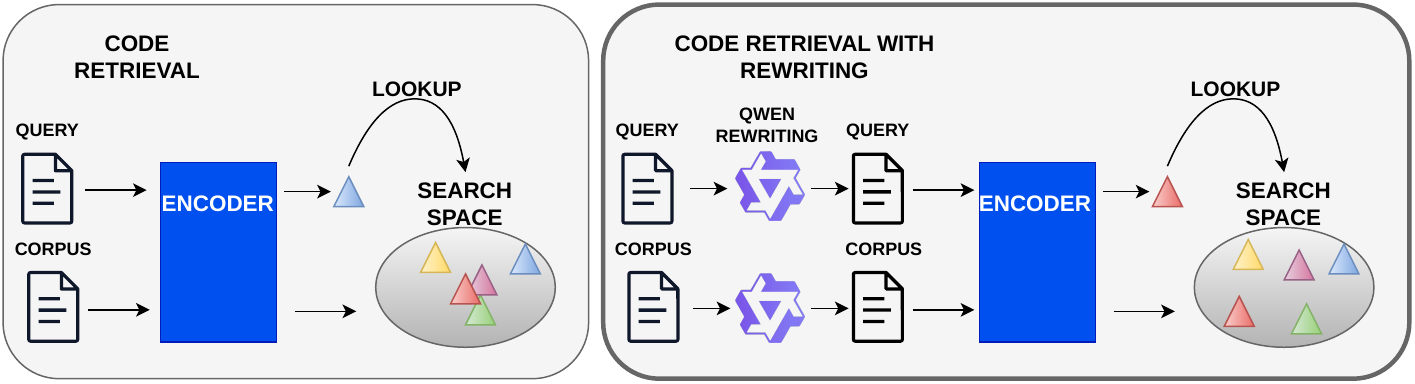}
  \caption{\textbf{Overview of the rewriting-augmented retrieval
    pipeline.} Queries and corpus documents are optionally passed
    through an LLM rewriter before being embedded by a frozen
    encoder. We study three rewriting strategies under two
    augmentation regimes: joint query--corpus (QC, online) and
    corpus-only (C, offline).}
  \label{fig:rewriting}
\end{figure}

Large Language Models (LLMs) have reshaped code retrieval, shifting
from lexical/AST-based methods to dense embedding-based
approaches~\citep{codebert}. However, current code encoders often
exhibit only a shallow understanding of program
behavior~\citep{graphcodebert}: they overweight surface-level
syntactic cues, mapping semantically distinct snippets to similar
vectors~\citep{semantic_relevance,unixcoder}. A recent line of work
addresses this by using LLMs to rewrite queries and corpora into a
more uniform form---either through stylistic
rephrasing~\citep{rewrite_by_normalizing} or a
code$\rightarrow$PseudoCode$\rightarrow$code
round-trip~\citep{pseudoBridge}. These approaches share two
limitations: (i) they operate at a \emph{single} representational
level (code), and (ii) they rewrite \emph{both} queries and corpus,
requiring an LLM call per query. Two questions follow:
\emph{how much representational shift actually helps}, and
\emph{when is the online LLM call worth it?}

We answer both through a systematic study varying two axes:
\emph{abstraction level} and \emph{online cost}. Using three
rewriters from independent model families
(Qwen3-Coder-30B, DeepSeek-Coder-V2-Lite-Instruct, Codestral-22B),
we instantiate three rewriting levels---(1) stylistic rephrasing
(following \citet{rewrite_by_normalizing}, our baseline),
(2) \emph{NL-enriched PseudoCode used directly as the retrieval
representation}, and (3) \emph{full natural-language transcription
used directly as the retrieval representation}. Levels (2) and (3)
are new retrieval representations: \citet{pseudoBridge} use
PseudoCode only as a transient bridge and ultimately retrieve over
code. Each level is evaluated under joint query--corpus (QC,
online) and corpus-only (C, offline) regimes. To explain
\emph{why} strategies work, we introduce two representation-level
diagnostics: the change in input token entropy $\Delta H$ (what the
encoder \emph{sees}) and the change in mean pairwise embedding
cosine $\Delta\bar{s}$ (how the encoder \emph{organizes} it).

Across six CoIR benchmarks (code-to-code, text-to-code, hybrid),
five encoders, and three rewriters, four findings emerge:
\textbf{(i) NL+QC is the strongest strategy for code-heavy
retrieval}, lifting MoSE-18 on CT-Contest from $0.23$ to $0.74$
NDCG@10 ($+0.51$ absolute) and remaining the best or tied-best
strategy on CT-Contest for all three rewriters.
\textbf{(ii) Corpus-only rewriting degrades retrieval in
${\sim}62\%$ of configurations} ($56/90$) relative to the unmodified baseline due to query--corpus
modality mismatch, while QC dominates C in $78/90$ paired comparisons.
\textbf{(iii) $\Delta H$ is a rewriter-agnostic predictor of
retrieval gain under QC} (Codestral: $\rho{=}{+}0.593$,
$p{<}0.001$; DeepSeek: $\rho{=}{+}0.274$; pooled non-Qwen:
$\rho{=}{+}0.436$, $p{<}0.001$; Qwen-only: $\rho{=}{+}0.356$).
\textbf{(iv) The best rewriting strategy is rewriter-dependent but
$\Delta H$ identifies it}: the strict Rephrase$<$Pseudo$<$NL
ordering is Qwen-specific, but $\Delta H$ correctly tracks the
best strategy per rewriter. All prompts, rewriting templates, and
experimental code will be released.

\section{Background and Related Work}
\label{subsec:related}

\paragraph{Code Information Retrieval (CIR).}
CIR retrieves software artifacts from a corpus in response to a
query, where both query and items may be code, text, or a hybrid
mixture. We use the \textsc{CoIR} benchmark
suite~\citep{coir}, which aggregates ten datasets across
text-to-code, code-to-code, and hybrid-code modalities and reports
NDCG@10 as the primary metric.

\paragraph{LLM-based rewriting for retrieval.}
\citet{gar} introduced Generation-Augmented Retrieval for
open-domain QA. \citet{rewrite_by_normalizing} extend this to code
by rephrasing snippets in the LLM's own writing style, normalizing
surface form---the current state of the art.
\citet{pseudoBridge} introduce a
code$\rightarrow$PseudoCode$\rightarrow$code round-trip in which
PseudoCode is used to align semantic content but is discarded
before retrieval.

\paragraph{What we add.}
These methods share four limitations: \textbf{(i) Representational
commitment}: each fixes a single abstraction level a priori (both
ultimately retrieve over code); no prior work evaluates PseudoCode
or snippet-level NL as the \emph{retrieval target}
(Table~\ref{tab:positioning}).
\textbf{(ii) Cost}: all require online LLM calls per query.
\textbf{(iii) Rewriter sensitivity}: prior work uses a single
rewriter, leaving generalization across families open.
\textbf{(iv) Diagnostics}: none characterize when rewriting is
worth the cost. We address all four: (a) treat PseudoCode and
snippet-level NL as direct retrieval representations; (b) unify all
three levels in a single controlled comparison; (c) add a
corpus-only variant; (d) evaluate across three independent
rewriter families; (e) provide a representation-level diagnostic
predictive of retrieval gain.

\paragraph{Scope of comparison with PseudoBridge.}
\citet{pseudoBridge} differs from our setup along three axes
simultaneously---two-step vs.\ single-step synthesis, fine-tuned
vs.\ frozen encoder, and code-level vs.\ rewritten-representation
retrieval---so a head-to-head would not isolate the effect we study.
We therefore include it in Table~\ref{tab:positioning} for taxonomic
completeness and use the single-axis rephrasing baseline of
\citet{rewrite_by_normalizing} as our controlled reference.

\begin{table}[t]
  \centering
  \footnotesize
  \setlength{\tabcolsep}{4pt}
  \renewcommand{\arraystretch}{1.1}
  \caption{\textbf{Positioning of our rewriting strategies relative to prior
    work.} Prior methods keep the retrieval target at the code level
    (optionally round-tripping through PseudoCode). We are the first to
    evaluate \textbf{NL-enriched PseudoCode} and \textbf{snippet-level full
    NL} as \emph{direct} retrieval representations. Rows in grey denote
    baselines. $\star$ denotes a retrieval representation not evaluated by
    prior work.}
  \label{tab:positioning}
  \begin{tabularx}{\linewidth}{@{}l *{3}{>{\centering\arraybackslash}X}@{}}
    \toprule
    \textbf{Method} & \textbf{Query form} & \textbf{Indexed form}
      & \textbf{Retrieval target} \\
    \midrule
    \rowcolor{gray!12}
    No rewriting \textit{(baseline)}
      & code / text & code & code \\
    \rowcolor{gray!12}
    Rephrasing~\citep{rewrite_by_normalizing} \textit{(baseline)}
      & code (rephr.) & code (rephr.) & code \\
    PseudoBridge~\citep{pseudoBridge}\footnotemark
      & code & code (pseudo.\ bridge) & code \\
    \midrule
    \textbf{Ours -- PseudoCode}~$\star$
      & \textbf{PseudoCode} & \textbf{PseudoCode}
      & \textbf{PseudoCode (snippet)} \\
    \textbf{Ours -- Natural Language}~$\star$
      & \textbf{NL} & \textbf{NL} & \textbf{NL (snippet)} \\
    \bottomrule
  \end{tabularx}
\end{table}

\footnotetext{PseudoBridge fine-tunes the encoder via a two-stage
contrastive pipeline over synthesised pseudo-code and style-augmented
code variants, and retrieves over code at inference; our single-step,
frozen-encoder setup makes \citet{rewrite_by_normalizing} the appropriate
single-axis baseline.}

\section{The Paraphrasing Strategy}
\label{sec:paraphrasing}

\begin{figure}[t]
\centering
\footnotesize

\begin{tcolorbox}[
    colback=gray!5,
    colframe=black!60,
    boxrule=0.5pt,
    arc=3pt,
    left=6pt,right=6pt,top=6pt,bottom=6pt
]

\noindent
\begin{minipage}[t]{0.235\linewidth}
\textbf{Original code}\par\smallskip
{\ttfamily
def first\_mis\_pos(num):\par
\quad seen = set(x for x in num\par
\quad\quad if x > 0)\par
\quad i = 1\par
\quad while i in seen:\par
\quad\quad i += 1\par
\quad return i\par
}
\end{minipage}
\hfill
\vrule
\hfill
\begin{minipage}[t]{0.235\linewidth}
\textbf{Rephrasing}\par\smallskip
{\ttfamily
def first\_mis\_pos(num):\par
\quad positives = \{v for v in num\par
\quad\quad if v > 0\}\par
\quad candidate = 1\par
\quad while candidate in positives:\par
\quad\quad candidate += 1\par
\quad return candidate\par
}
\end{minipage}
\hfill
\vrule
\hfill
\begin{minipage}[t]{0.235\linewidth}
\textbf{PseudoCode}\par\smallskip
{\ttfamily
FUNCTION
first\_mis\_pos(num):\par
\quad keep only positive numbers\par
\quad store them for fast lookup\par
\quad candidate $\leftarrow$ 1\par
\quad WHILE candidate exists:\par
\quad\quad candidate $\leftarrow$ candidate + 1\par
\quad RETURN candidate\par
}
\end{minipage}
\hfill
\vrule
\hfill
\begin{minipage}[t]{0.235\linewidth}
\textbf{Natural Language}\par\smallskip
Return the smallest missing positive integer by ignoring non-positive values and checking integers from 1 upward.
\end{minipage}

\end{tcolorbox}

\caption{\textbf{Example of the rewriting hierarchy.}
A function is transformed from its original implementation
(\emph{Original code}) to a stylistically normalized version
(\citet{rewrite_by_normalizing} \emph{Rephrasing}), then to NL-enriched PseudoCode
(\emph{PseudoCode}), and finally to a full natural-language
description (\emph{Natural Language}).  In our pipeline, the
PseudoCode and Natural language forms are used \emph{directly} as
the retrieval representation.}

\label{fig:rewriting_hierarchy}
\end{figure}

Prior work explores stylistic normalization through code
rephrasing~\citep{rewrite_by_normalizing,pseudoBridge}. We
hypothesize that alternative representations---\textbf{natural
language descriptions} and \textbf{NL-enriched PseudoCode}---used
as the sole code representation can yield superior retrieval
performance. We also investigate the efficiency limitation of
state-of-the-art methods, which require an LLM call \emph{per
query} (\textsc{QC-manipulation}); we ask whether rewriting only
the corpus once offline (\textsc{C-manipulation}) is empirically
viable.

\paragraph{Two new retrieval representations.}
We introduce \textbf{NL-enriched PseudoCode} and
\textbf{snippet-level Natural Language} as \emph{direct} retrieval
targets (\Cref{fig:rewriting_hierarchy}). Unlike
\citet{pseudoBridge}, who use pseudo code as a transient bridge in
a code$\rightarrow$pseudo$\rightarrow$code pipeline, we treat
PseudoCode (resp.\ NL) as the final representation passed to the
encoder. The rewriter is prompted to first comprehend the snippet
and then generate the target representation; the same form is used
both to index documents and to encode queries. For text-to-code
tasks under QC, the LLM generates the target form directly from the
NL request (Rephrase: code; Pseudo: commented PseudoCode; NL:
restyled NL). Positioning relative to prior work is summarized in
Table~\ref{tab:positioning}.

\paragraph{Baselines.}
We compare against (i)~the unmodified corpus and queries, and
(ii)~the stylistic rephrasing of \citet{rewrite_by_normalizing}.

\paragraph{Evaluation setup.}
We evaluate on six CoIR test sets~\citep{coir}: codetrans-contest,
codetrans-dl (code-to-code); apps, cosqa (text-to-code);
StackOverflow-QA, CodeFeedback-MT (hybrid). We select six of the ten CoIR tasks to span all three task families (code-to-code, text-to-code, hybrid) while keeping the full 5 encoders × 3 strategies  × 6 benchmarks = 90 configurations tractable within our compute budget (\S\ref{subsec:repr-results}).  All results use NDCG@10 metric.
Each strategy is evaluated under the same prompt family,
and rewriter for a controlled comparison, on general-purpose
encoders (\textsc{Qwen3-Emb}~\citep{qwen3emb},
\textsc{E5-Base-V2}~\citep{e5}) and code-specialized ones
(\textsc{MoSE-18}~\citep{mose}, \textsc{CodeXEmbed}~\citep{codexembed},
\textsc{UniXCoder}~\citep{unixcoder}). The main rewriter is
\textsc{Qwen3-Coder-30B-A3B-Instruct}~\cite{qwen3};
\S\ref{subsec:cross-rewriter} additionally evaluates
\textsc{DeepSeek-Coder-V2-Lite-Instruct} (16B MoE) and
\textsc{Codestral-22B} (Mistral, 22B dense) to rule out
rewriter-specific artifacts.

\subsection{Representational Analysis of Rewriting Effects}
\label{sec:repr-analysis}

To understand \emph{how} rewriting strategies yield different
outcomes, we analyze input- and embedding-level corpus properties
via two complementary diagnostics, computed on both baseline and
rewritten corpora under identical batching.

\paragraph{Input token entropy.}
For all non-padding tokens in a batch, we compute the Shannon
entropy $H = -\sum_{v \in \mathcal{V}} \hat{p}(v)\log_2 \hat{p}(v)$
of the empirical token-frequency distribution. This captures the
\emph{lexical diversity} the encoder receives: code-heavy text
concentrates mass on a small set of syntactic tokens (low entropy),
whereas NL-rich text spreads mass across a broader vocabulary. We
report
$\Delta H = H_{\text{rewritten}} - H_{\text{baseline}}$.

\paragraph{Embedding pairwise cosine similarity.}
For $\ell_2$-normalized embeddings $\{\mathbf{e}_i\}_{i=1}^{B}$ we
compute the mean off-diagonal cosine
$\bar{s} = \frac{1}{B(B-1)} \sum_{i \neq j} \mathbf{e}_i^{\top}\mathbf{e}_j$,
a measure of representation isotropy: lower values indicate more
discriminative spread; higher values indicate anisotropic collapse.
We report
$\Delta\bar{s} = \bar{s}_{\text{rewritten}} - \bar{s}_{\text{baseline}}$.
Together $\Delta H$ and $\Delta\bar{s}$ disentangle tokenizer-level
distributional shifts from embedding-level geometric changes.

%% ===========================================================================
%% ===========================================================================
%% ===========================================================================
\section{Main Evaluation}
\label{sec:main-eval}

\begin{figure}[!ht]
    \centering
    \includegraphics[width=\linewidth]{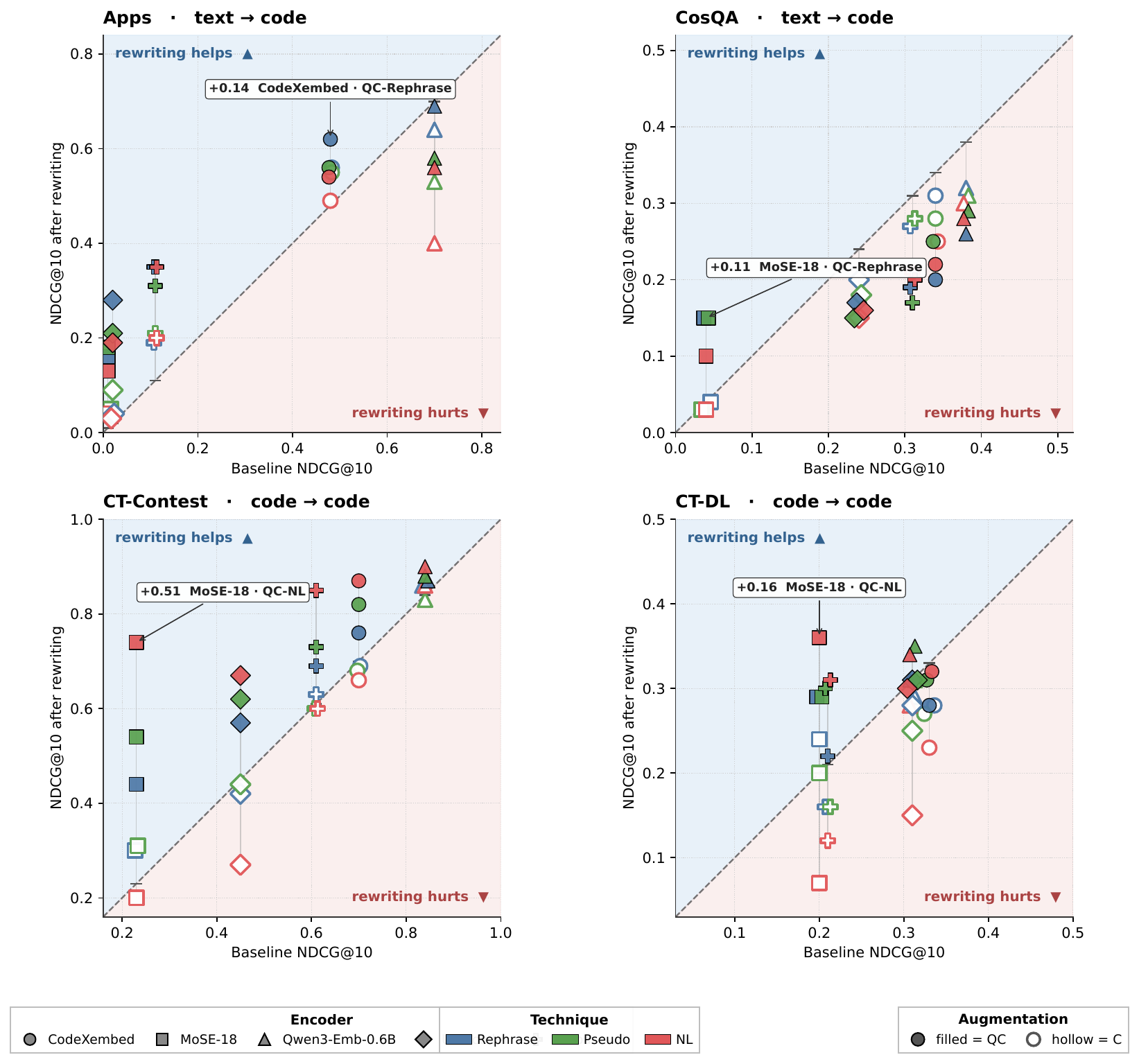}
\caption{\textbf{Per-task NDCG@10 retrieval performances.} Representation after rewriting compared to the original baseline for the five encoders under query+corpus (QC, filled markers) and corpus-only (C, hollow markers) augmentation. Marker shape denotes the encoder, and color indicates the technique (Rephrase / Pseudo / NL); six variants are stacked vertically above each encoder’s baseline. Annotations highlight the largest QC improvement for each task. QC–NL is most effective for smaller encoders on code-intensive tasks, while C consistently underperforms QC.}
    \label{fig:rewriting_results}
\end{figure}

We evaluate on six CoIR tasks spanning three families:
\emph{code-to-code} (CT-Contest, CT-DL), \emph{text-to-code}
(Apps, CosQA), and \emph{hybrid} (StackOverflow-QA,
CodeFeedback-MT). Full per-cell NDCG@10 appears in Appendix
Tables~\ref{table:resultst2c}--\ref{table:resultshyb}.

\paragraph{Code-to-code.}
QC-NL is the best strategy for every encoder on CT-Contest and for
four of five on CT-DL (see Figure~\ref{fig:rewriting_results}), with gains scaling inversely with encoder
capacity: MoSE-18 improves by $+0.51$ NDCG@10 on CT-Contest
($0.23\!\to\!0.74$) and $+0.16$ on CT-DL; E5-base-v2 by $+0.24$ and
$+0.10$. PseudoCode sits monotonically between Rephrasing and NL.
CodeXEmbed on CT-DL (baseline $0.33\!>\!$all rewrites) indicates
that sufficiently strong code encoders saturate the benefit.

\paragraph{Text-to-code.}
The hierarchy breaks down once queries are already in natural language (Figure~\ref{fig:rewriting_results}).
On Apps, QC-Rephrasing is the best \emph{average} configuration
(CodeXEmbed $+0.14$); on CosQA, \emph{no} QC configuration improves
over the strongest baselines (Qwen3-Emb $0.38$, CodeXEmbed $0.34$):
translating already-NL queries against an NL-rewritten corpus
erases residual syntactic signal without creating new alignment.

\paragraph{Hybrid (Table~\ref{table:resultshyb_aggregated}).}
The three strategies collapse to within $0.01$ NDCG@10 under QC,
with PseudoCode and NL tied at $0.26$ average. Gains come almost
entirely from CodeFeedback-MT ($0.07\!\to\!0.10$, $+43\%$ rel.).
C-NL is the only configuration that drops below baseline on average.

\begin{table}[!ht]
\small
\caption{\textbf{Hybrid retrieval, NDCG@10 aggregated across five encoders.}
  Under QC, PseudoCode and NL tie for best; C-NL is the only setting
  below the unmodified baseline on average.}
\centering
\begin{tabular}{llcc|c}
\toprule
Technique & Aug. & StackOverflow-QA & CodeFeedback-MT & Avg. \\
\midrule
Rephrasing & QC & 0.40 & 0.10 & 0.25 \\
Rephrasing & C  & 0.38 & 0.08 & 0.23 \\
\rowcolor{gray!12}
PseudoCode & QC & 0.42 & 0.10 & \textbf{0.26} \\
PseudoCode & C  & 0.39 & 0.08 & 0.24 \\
\rowcolor{gray!12}
NL         & QC & 0.42 & 0.10 & \textbf{0.26} \\
NL         & C  & 0.34 & 0.07 & 0.21 \\
\midrule
Baseline   & -- & 0.41 & 0.07 & 0.24 \\
\bottomrule
\end{tabular}
\label{table:resultshyb_aggregated}
\end{table}

\paragraph{Three cross-cutting patterns.}
(i)~\textbf{QC dominates C in 78/90 paired configurations ($86.7\%$):}
C-NL degrades MoSE-18's average from $0.12$ to $0.08$; joint
rewriting is necessary to prevent query--corpus modality mismatch and to gain stylistic normalization.
(ii)~\textbf{Gains scale inversely with encoder strength}: averaged
over four pure retrieval tasks, QC-NL lifts MoSE-18 by $+175\%$
rel. ($0.12\!\to\!0.33$), UniXcoder by $+32\%$, E5-base-v2 by
$+39\%$, but leaves Qwen3-Emb-0.6B flat or slightly worse
($0.56\!\to\!0.52$); rewriting is most valuable as a
\emph{remediation layer} for lightweight encoders.
(iii)~\textbf{Abstraction value decays with query NL content}:
Rephrase$<$Pseudo$<$NL holds on code-to-code, becomes inconsistent
on text-to-code, and collapses to within $0.01$ on hybrid.

\paragraph{Effect of rewriter size.}
A separate in-vitro study on CT-Contest using the
\textsc{Qwen2.5-Coder-Instruct} family (1.5B--14B,
Appendix Table~\ref{table:in_vitro}) shows larger rewriters generally improve retrieval quality on average, but the trend is not monotonic for every encoder–strategy pair; gains are linked to rewriter quality, with
corresponding hardware/latency constraints for practitioners.

%% -------------------------------------------------------------------
\section{Representational Analysis}
\label{subsec:repr-results}

\paragraph{Scope of the representational analysis.}
We restrict the diagnostic analysis to the four pure code-to-code
and text-to-code benchmarks: hybrid corpora already mix prose and
code in variable proportions, so their baseline entropy and embedding
geometry reflects the intrinsic NL/code ratio rather than the
rewriting-induced shift we aim to measure.  Hybrid benchmarks
instead serve as an external validity check at the retrieval level. Tables~\ref{tab:repr} and~\ref{tab:tokenisation} characterize how
each rewriting strategy reshapes the tokenizer-level and
encoder-level properties of the corpus.

% \begin{table}[t]
%   \centering
%   \caption{Mean change in input token entropy ($\Delta H$, bits) and
%     embedding pairwise cosine ($\Delta\bar{s}$) after corpus rewriting,
%     averaged across the four evaluation tasks.  Arrows indicate the
%     direction typically associated with improved retrieval
%     ($\uparrow$\,for\,$\Delta H$, $\downarrow$\,for\,$\Delta\bar{s}$).}
%   \label{tab:repr}
%   \small
%   \begin{tabular}{ll rr}
%     \toprule
%     \textbf{Encoder} & \textbf{Technique}
%       & $\Delta H\!\uparrow$ & $\Delta\bar{s}\!\downarrow$ \\
%     \midrule
%     \multirow{3}{*}{\textsc{CodexEmbed}}
%       & Rephrase      & $+0.79$ & $-0.031$ \\
%       & PseudoCode    & $+1.15$ & $-0.005$ \\
%       & NL            & $+1.42$ & $-0.082$ \\
%     \midrule
%     \multirow{3}{*}{\textsc{MoSE-18}}
%       & Rephrase      & $+0.60$ & $-0.057$ \\
%       & PseudoCode    & $+0.95$ & $-0.031$ \\
%       & NL            & $+0.90$ & $-0.068$ \\
%     \midrule
%     \multirow{3}{*}{\textsc{Qwen3-Emb}}
%       & Rephrase      & $+0.55$ & $-0.133$ \\
%       & PseudoCode    & $+0.82$ & $-0.128$ \\
%       & NL            & $+0.47$ & $-0.131$ \\
%     \midrule
%     \multirow{3}{*}{\textsc{UniXcoder}}
%       & Rephrase      & $+0.60$ & $-0.04$ \\
%       & PseudoCode    & $+1.06$ & $+0.08$ \\
%       & NL            & $+1.07$ & $-0.15$ \\
%     \midrule
%     \multirow{3}{*}{\textsc{E5-base-v2}}
%       & Rephrase      & $+0.80$ & $+0.001$ \\
%       & PseudoCode    & $+1.16$ & $+0.016$ \\
%       & NL            & $+1.44$ & $-0.018$ \\
%     \bottomrule
%   \end{tabular}
% \end{table}

\begin{table*}[t]
  \centering
  \caption{\textbf{Mean change in input token entropy ($\Delta H$, bits) and
    embedding pairwise cosine ($\Delta\bar{s}$).} Results are reported after corpus rewriting,
    averaged across the four evaluation tasks.  Arrows indicate the
    direction typically associated with improved retrieval
    ($\uparrow$\,for\,$\Delta H$, $\downarrow$\,for\,$\Delta\bar{s}$).}
  \label{tab:repr}
  \small
  \setlength{\tabcolsep}{5pt}
  \resizebox{\linewidth}{!}{%
  \begin{tabular}{ll cc cc cc cc cc}
    \toprule
    & & \multicolumn{2}{c}{\textsc{CodexEmbed}}
      & \multicolumn{2}{c}{\textsc{MoSE-18}}
      & \multicolumn{2}{c}{\textsc{Qwen3-Emb}}
      & \multicolumn{2}{c}{\textsc{UniXcoder}}
      & \multicolumn{2}{c}{\textsc{E5-base-v2}} \\
    \cmidrule(lr){3-4}\cmidrule(lr){5-6}\cmidrule(lr){7-8}\cmidrule(lr){9-10}\cmidrule(lr){11-12}
    \textbf{Technique} &
      & $\Delta H\!\uparrow$ & $\Delta\bar{s}\!\downarrow$
      & $\Delta H\!\uparrow$ & $\Delta\bar{s}\!\downarrow$
      & $\Delta H\!\uparrow$ & $\Delta\bar{s}\!\downarrow$
      & $\Delta H\!\uparrow$ & $\Delta\bar{s}\!\downarrow$
      & $\Delta H\!\uparrow$ & $\Delta\bar{s}\!\downarrow$ \\
    \midrule
    Rephrase   & & $+0.79$ & $-0.031$ & $+0.60$ & $-0.057$ & $+0.55$ & $-0.133$ & $+0.60$ & $-0.04\phantom{0}$ & $+0.80$ & $+0.001$ \\
    PseudoCode & & $+1.15$ & $-0.005$ & $+0.95$ & $-0.031$ & $+0.82$ & $-0.128$ & $+1.06$ & $+0.08\phantom{0}$ & $+1.16$ & $+0.016$ \\
    NL         & & $+1.42$ & $-0.082$ & $+0.90$ & $-0.068$ & $+0.47$ & $-0.131$ & $+1.07$ & $-0.15\phantom{0}$ & $+1.44$ & $-0.018$ \\
    \bottomrule
  \end{tabular}}
\end{table*}

\begin{figure*}[t]
  \centering
  \includegraphics[width=\linewidth]{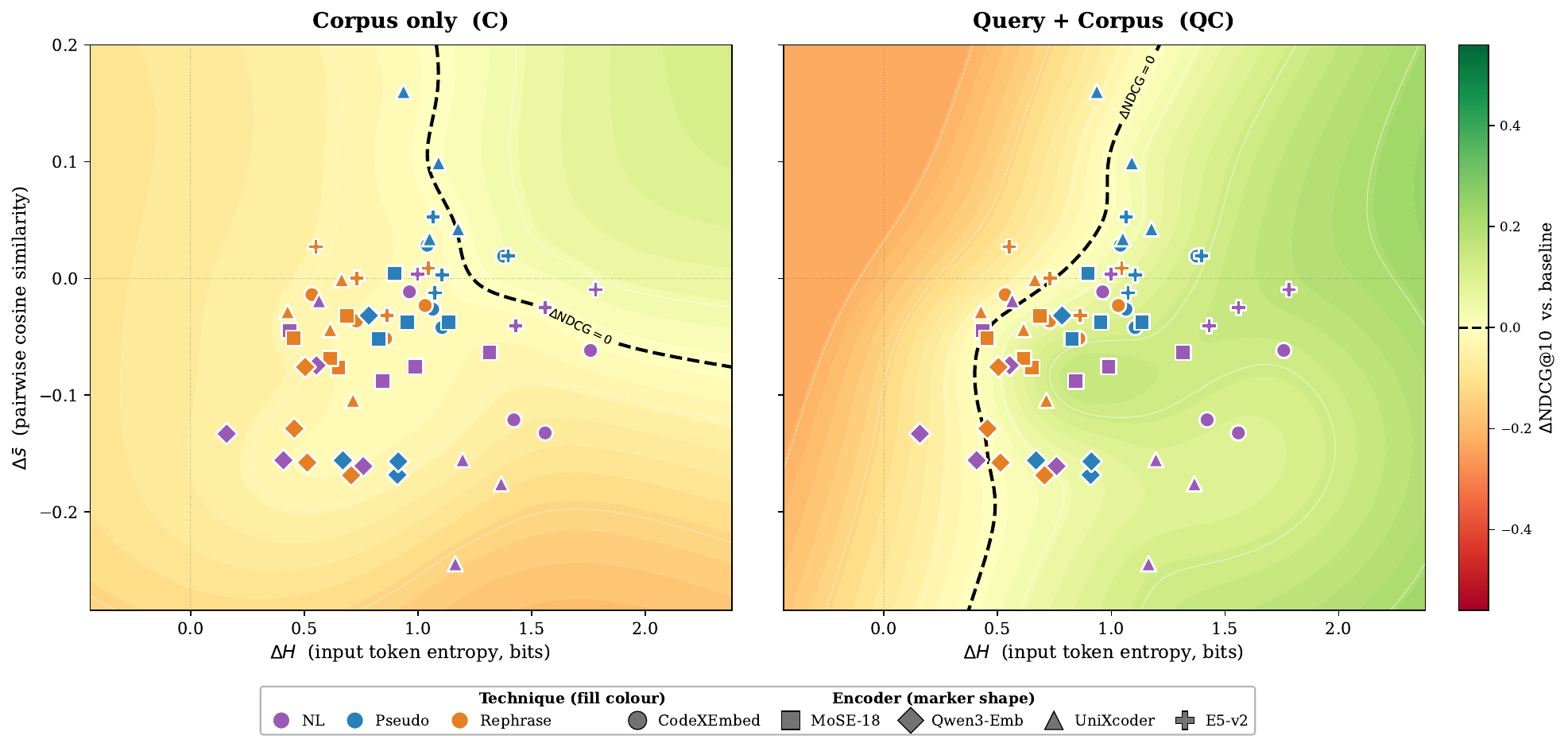}
  \caption{\textbf{Retrieval efficacy landscape in representational-shift space.}
Each point is an (encoder, task, technique) configuration at $(\Delta H,\,\Delta\bar{s})$. The background shows $\Delta\text{NDCG@10}$ relative to the unmodified baseline, interpolated with a thin-plate-spline RBF; white contours are iso-$\Delta\text{NDCG}$, and the dashed black line is $\Delta\text{NDCG}=0$.
\textbf{Left:} corpus-only (C)—large representational shifts enter the red zone, where retrieval worsens if the query is unchanged.
\textbf{Right:} query\,+\,corpus (QC)—the same points move to green, indicating that co-transforming the query recovers and often exceeds baseline performance.
Marker fill denotes rewriting technique (\legendswatch{violet}~NL, \legendswatch{blue}~Pseudo, \legendswatch{orange}~Rephrase); marker shape denotes encoder.}
  \label{fig:landscape}
\end{figure*}

\begin{figure*}[!ht]
  \centering
  \includegraphics[width=\linewidth]{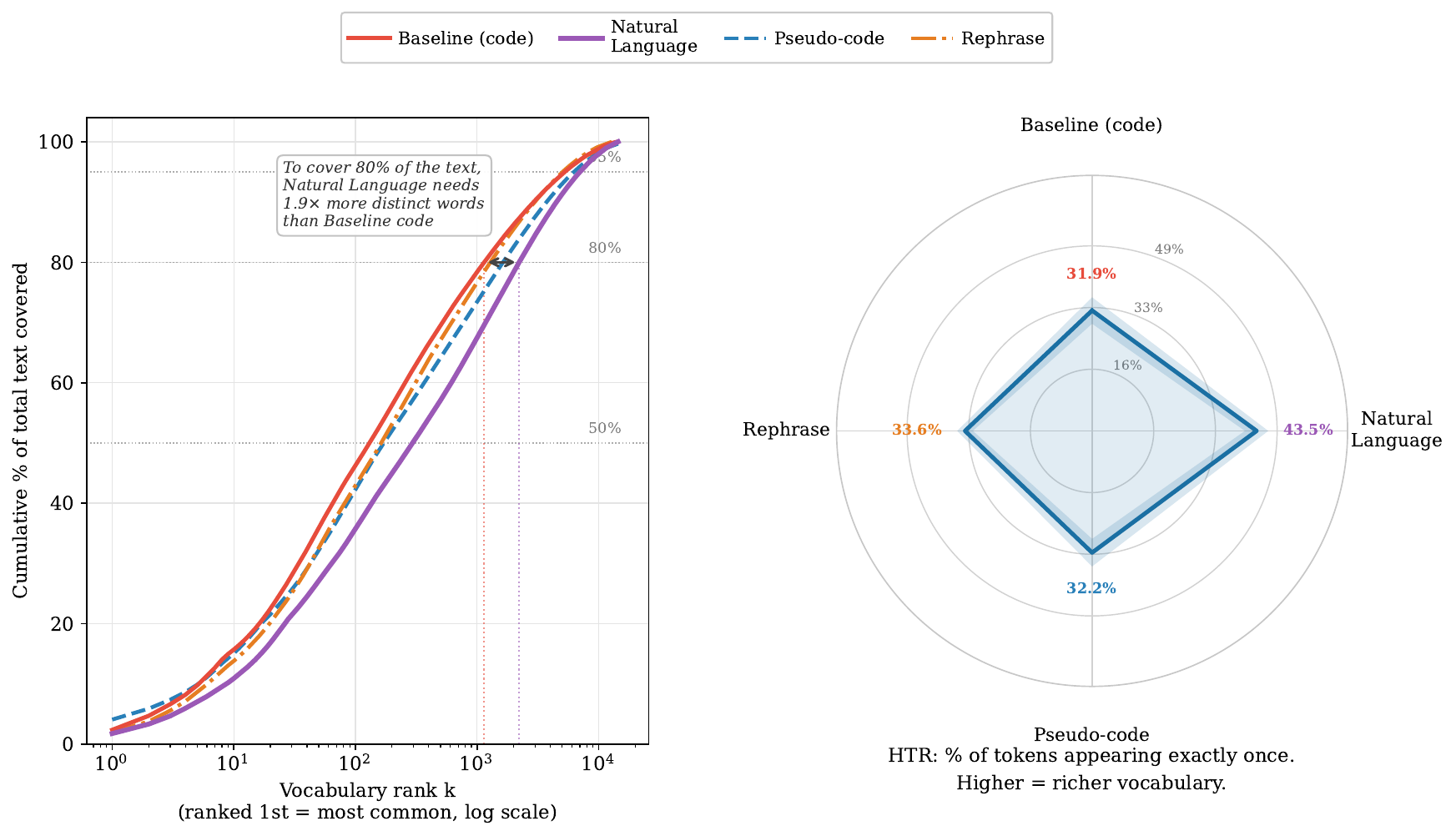}
  \caption{\textbf{Vocabulary coverage and lexical-richness
    diagnostics.}  \textbf{Left:} cumulative fraction of total
    tokens covered by the top-$k$ vocabulary ranks (log scale).
    Natural-language rewriting requires ${\sim}1.9\times$ more
    distinct words than the code baseline to reach $80\%$ coverage,
    confirming a flatter token distribution.
    \textbf{Right:} radar plot of the hapax rate (Hapax\%, the
    fraction of tokens appearing exactly once) across the four
    representations.  NL achieves the highest hapax rate ($43.5\%$),
    indicating the richest long-tail vocabulary.}
  \label{fig:cdf_radar}
\end{figure*}

\paragraph{Token entropy increases monotonically with abstraction.}
For four of five encoders,
$\Delta H_{\text{Rephrase}} < \Delta H_{\text{Pseudo}} < \Delta H_{\text{NL}}$
holds (Table~\ref{tab:repr}); Qwen3-Emb is the exception, since its
151k-token vocabulary absorbs NL diversity into subword merges. The
largest gains accrue to small-vocabulary encoders (CodeXEmbed,
E5-base-v2: $\Delta H{\approx}{+}1.4$ bits under NL---roughly
doubling the effective alphabet). NL also yields the richest tail:
Hapax\% reaches $47.5\%$ (Qwen3-Emb) and $45.4\%$ (MoSE-18) vs.\
baselines of $36.6\%$ and $33.3\%$
(Table~\ref{tab:tokenisation}); Top-20\% mass drops by up to
$21$\,pp. PseudoCode maximizes raw unique types but its Hapax\%
stays near baseline (with many quasi-syntactic tokens recurring).
Figure~\ref{fig:cdf_radar} corroborates this: NL requires
${\sim}1.9{\times}$ more distinct words than code to cover $80\%$
of the text and achieves the highest overall Hapax ($43.5\%$).

\paragraph{Embedding isotropy improves under NL rewriting.}
NL reduces mean pairwise cosine for all five encoders
($\Delta\bar{s} \leq -0.018$), most strongly for UniXcoder
($-0.15$) and Qwen3-Emb ($-0.131$). PseudoCode is the least
consistent, increasing $\Delta\bar{s}$ for UniXcoder ($+0.08$) and
E5-base-v2 ($+0.016$): residual syntactic structure can push
representations closer for some encoders.

\paragraph{Retrieval efficacy landscape.}
Figure~\ref{fig:landscape} projects every (encoder, task, technique)
configuration into $(\Delta H,\,\Delta\bar{s})$ space with
$\Delta\text{NDCG@10}$ as the background surface. Under C (left),
configurations with large representational shifts occupy the red
zone; the $\Delta\text{NDCG}{=}0$ contour runs diagonally,
indicating that any substantial corpus transformation without a
matching query transformation pushes retrieval below baseline.
Under QC (right), the same points migrate into the green zone---NL
points for MoSE-18 and E5-base-v2 land in the darkest region.

\paragraph{Correlation analysis.}
Table~\ref{tab:correlations} quantifies the visual pattern. Under
QC, $\Delta H$ is the sole significant predictor of retrieval gain
($\rho{=}{+}0.356$, $p{<}0.01$; $r{=}{+}0.319$, $p{<}0.05$);
$\Delta\bar{s}$ shows no significant association
($\rho{=}{-}0.064$). Under C, neither metric reaches significance,
where modality mismatch and missing query-side normalization
dominate. The two diagnostics are largely independent
($\rho{=}{+}0.229$, $p{=}0.078$), capturing complementary aspects.
Per \S\ref{subsec:cross-rewriter}, the QC correlation replicates
across DeepSeek and Codestral.

\paragraph{Efficiency.}
On an H100-80GB serving \textsc{Qwen3-Coder-30B-A3B-Instruct} (FP16,
vLLM, 512-token context, ${\sim}115$\,tok/s), rewriting the four
CoIR corpora (${\sim}38$K snippets) takes ${\sim}16.5$ GPU-hours
(NL) / ${\sim}11$ (Rephrasing) as a one-time offline cost; QC adds
${\sim}725$\,ms of decoding latency per query. Combined with the
above results, this yields a deployment decision framework: use QC
rewriting as a remediation layer when a lightweight encoder is
deployed on code-dominant queries, and skip it when a strong
encoder or NL-rich query is available.

\begin{table}[t]
\centering
\caption{% 
    \textbf{Correlation table between $\Delta H$, $\Delta\bar{s}$ and $\Delta$\,NDCG@10.}
  Spearman and Pearson correlations between representational-shift
  diagnostics ($\Delta H$: token entropy change; $\Delta\bar{s}$:
  embedding cosine similarity change) and retrieval gain
  ($\Delta$\,NDCG@10), across $n{=}60$ encoder--task--technique
  configurations under corpus-only~(C) and query+corpus~(QC) settings.
  $\Delta H$ is the sole significant predictor of retrieval gain, and
  only in the QC setting; the two diagnostics are largely independent
  ($\rho{=}{+}0.229$, $p{=}0.078$).  The QC correlation replicates
  across two independent rewriter families
  (Table~\ref{tab:cross-rewriter-correlation}).  Two-sided $p$-values;
  $^{*}p{<}0.05$, $^{**}p{<}0.01$.}
\label{tab:correlations}
\small
\setlength{\tabcolsep}{6pt}
\begin{tabular}{lcccc}
\toprule
 & \multicolumn{2}{c}{\textbf{C (corpus only)}}
 & \multicolumn{2}{c}{\textbf{QC (query+corpus)}} \\
\cmidrule(lr){2-3}\cmidrule(lr){4-5}
\textbf{Pair} ($n{=}60$)
  & Spearman $\rho$ & Pearson $r$
  & Spearman $\rho$ & Pearson $r$ \\
\midrule
$\Delta H$      \ vs\ $\Delta\text{NDCG@10}$
  & $+0.020$
  & $+0.086$
  & $\mathbf{+0.356^{**}}$
  & $\mathbf{+0.319^{*}}$ \\
$\Delta\bar{s}$ \ vs\ $\Delta\text{NDCG@10}$
  & $+0.108$
  & $+0.222$
  & $-0.064$
  & $-0.081$ \\
\midrule
$\Delta H$      \ vs\ $\Delta\bar{s}$
  & \multicolumn{2}{c}{$\rho = {+0.229}\ (p = 0.078)$}
  & \multicolumn{2}{c}{$r = {+0.146}\ (p = 0.267)$} \\
\bottomrule
\end{tabular}
\end{table}

%% -------------------------------------------------------------------
%% CROSS-REWRITER ROBUSTNESS
%% -------------------------------------------------------------------
\section{Cross-Rewriter Robustness}
\label{subsec:cross-rewriter}

To check whether conclusions are rewriter-specific, we replicate
the core experiments with two additional rewriters from independent
model families:
\textsc{DeepSeek-Coder-V2-Lite-Instruct} (16B MoE, ${\sim}2.4$B active)
and \textsc{Codestral-22B} (Mistral, 22B dense), on the two CoIR
tasks that most sharply discriminate among strategies (CT-Contest
and CosQA).

\paragraph{NL rewriting generalizes; strategy ordering is rewriter-dependent.}
On CT-Contest (Table~\ref{tab:cross-rewriter-aggregate}), NL rewriting is best for Qwen and DeepSeek (Qwen $0.81$, DeepSeek $0.65$) and competitive for Codestral (Codestral $0.72$), while the best strategy is rewriter-dependent. The strict
Rephrase$<$Pseudo$<$NL ordering does not replicate uniformly:
Codestral-Rephrase reaches $0.74$ (its best), and DeepSeek-Pseudo
underperforms DeepSeek-Rephrase. The advantage of NL rewriting is
a property of the task, while Rephrase vs.\ Pseudo ranking is a
property of the rewriter. Per-encoder numbers appear in Appendix
Tables~\ref{table:cross_contest}--\ref{table:cross_cosqa}.

\begin{table}[t]
  \centering
  \small
  \caption{\textbf{Multiple rewriter retrieval performances.} Mean NDCG@10 across five encoders per (rewriter, strategy)
    on two contrasting CoIR tasks.  \textbf{Bold} marks the best
    strategy per rewriter per task.  NL remains the best or
    tied-best strategy for all three rewriters on the code-heavy
    CT-Contest task; on the NL-heavy CosQA task, no rewriting
    strategy beats the baseline for any rewriter, confirming that
    rewriting's failure on NL-heavy queries is an intrinsic property
    of the task, not a rewriter artifact.}
  \label{tab:cross-rewriter-aggregate}
  \setlength{\tabcolsep}{6pt}
  \renewcommand{\arraystretch}{1.1}
  \begin{tabular}{lcccc}
    \toprule
    \textbf{Rewriter} & \textbf{Family} & \textbf{Rephrase} & \textbf{Pseudo} & \textbf{NL} \\
    \midrule
    \multicolumn{5}{l}{\emph{CT-Contest (code-to-code)}} \\
    \midrule
    Qwen3-Coder-30B       & Qwen      & 0.67 & 0.72 & \textbf{0.81} \\
    DeepSeek-V2-Lite      & DeepSeek  & 0.58 & 0.53 & \textbf{0.65} \\
    Codestral-22B         & Mistral   & \textbf{0.74} & 0.66 & 0.72 \\
    \midrule
    \multicolumn{5}{l}{\emph{CosQA (text-to-code)}} \\
    \midrule
    Qwen3-Coder-30B       & Qwen      & 0.21 & 0.22 & 0.20 \\
    DeepSeek-V2-Lite      & DeepSeek  & 0.17 & 0.13 & 0.16 \\
    Codestral-22B         & Mistral   & 0.16 & 0.17 & 0.17 \\
    \bottomrule
  \end{tabular}
\end{table}

\paragraph{The $\Delta H$ diagnostic replicates across rewriter families.}
We recompute the $(\Delta H, \Delta\text{NDCG@10})$ correlation
per rewriter ($n{=}30$: 5 encoders $\times$ 3 strategies $\times$
2 tasks) and pooled (Table~\ref{tab:cross-rewriter-correlation}).
The correlation replicates with \emph{stronger} magnitude on
Codestral ($\rho{=}{+}0.593$, $p{<}0.001$) than on Qwen, preserves
sign on DeepSeek ($\rho{=}{+}0.274$), and reaches
$\rho{=}{+}0.436$, $p{<}0.001$ when pooled across non-Qwen
rewriters. $\Delta H$ is therefore a rewriter-agnostic predictor.

\begin{table}[t]
  \centering
  \small
  \caption{\textbf{Cross-rewriter replication of the $(\Delta H,\,
    \Delta\text{NDCG@10})$ correlation under QC.}  Independent
    experiments on DeepSeek and Codestral reproduce the positive
    correlation observed in our original Qwen analysis; the pooled
    non-Qwen correlation is \emph{stronger} than the Qwen-only result,
    indicating that $\Delta H$ captures a retrieval-relevant property
    that is not rewriter-specific.  Two-sided $p$-values;
    $^{*}p{<}0.05$, $^{**}p{<}0.01$, $^{***}p{<}0.001$.}
  \label{tab:cross-rewriter-correlation}
  \setlength{\tabcolsep}{6pt}
  \begin{tabular}{llrrr}
    \toprule
    \textbf{Rewriter} & \textbf{Family} & $n$ & \textbf{Spearman $\rho$} & \textbf{Pearson $r$} \\
    \midrule
    Qwen3-Coder-30B (original)      & Qwen      & 60 & $+0.356^{**}$  & $+0.319^{*}$ \\
    DeepSeek-Coder-V2-Lite          & DeepSeek  & 30 & $+0.274$       & $+0.292$ \\
    Codestral-22B                   & Mistral   & 30 & $+0.593^{***}$ & $+0.529^{**}$ \\
    \midrule
    \textbf{Pooled (DeepSeek + Codestral)} & \textbf{--} & \textbf{60} & $\mathbf{+0.436^{***}}$ & $\mathbf{+0.432^{***}}$ \\
    \bottomrule
  \end{tabular}
\end{table}

\paragraph{$\Delta H$ identifies the best strategy per rewriter, bidirectionally.}
Because the best strategy differs across rewriters
(Table~\ref{tab:cross-rewriter-aggregate}) and $\Delta H$ correlates
with retrieval gain \emph{within} each rewriter, practitioners can
use $\Delta H$ to select the right strategy without running full
retrieval evaluation. The diagnostic also operates bidirectionally:
on NL-heavy CosQA, both new rewriters yield small or negative mean
$\Delta H$ (DeepSeek: $-0.11$; Codestral: $-0.36$) vs.\ CT-Contest
($+0.39$, $+0.47$), and correspondingly $\Delta\text{NDCG}$ is
uniformly negative on CosQA for all three rewriters.

\section{Conclusion}
\label{sec:conclusion}

We introduced two new retrieval representations---NL-enriched
PseudoCode and snippet-level full Natural Language---and placed
them, with the rephrasing baseline of
\citet{rewrite_by_normalizing}, in a controlled abstraction
hierarchy evaluated across six CoIR benchmarks, five encoders, and
three rewriter families. Four findings reframe rewriting as a
cost--benefit decision:
(i) NL+QC yields the largest gains (up to $+0.51$ NDCG@10 on CT-Contest
for MoSE-18), especially for lightweight encoders, and is best or
competitive on code-to-code tasks across all three rewriters;
(ii) corpus-only rewriting degrades retrieval in ${\sim}62\%$ of
configurations, while QC outperforms C in $78/90$ paired comparisons;
(iii) $\Delta H$ is a significant cross-rewriter predictor of
retrieval gain under QC (pooled non-Qwen $\rho{=}{+}0.436$,
$p{<}0.001$);
(iv) the best strategy is rewriter-dependent but $\Delta H$
identifies it. Practitioners should deploy QC rewriting as a
remediation layer for small encoders on code-dominant queries, use
$\Delta H$ for strategy selection, and skip rewriting when a strong
encoder or NL-rich query is available.

\section{Limitations and Broader Impact}
\label{sec:limitations}
\label{sec:impact}

\paragraph{Limitations} Our study has four limitations.  \textbf{(i) Rewriter coverage.}
While our cross-rewriter analysis (\S\ref{subsec:cross-rewriter})
spans three independent model families (Qwen, DeepSeek, Mistral) and
our size-effect analysis (Appendix Table~\ref{table:in_vitro})
covers four scales within the Qwen family, we do not evaluate closed-source
rewriters (e.g., GPT-4o, Claude); extending the correlation study
to frontier proprietary models is an open direction.
\textbf{(ii) Language coverage.}  CoIR spans multiple languages but
is Python-heavy; behavior on low-resource languages remains open.
\textbf{(iii) Diagnostic scope.}  $\Delta H$ and $\Delta\bar{s}$ are
corpus-level aggregates and do not predict \emph{per-query} gains;
extending them to per-query confidence estimation is an open
direction.  \textbf{(iv) Deployment assumptions.}  Our latency
measurements assume a single H100 without production-grade batching,
caching, or query-side pre-computation; these optimizations could
further shift the QC vs.\ C trade-off toward QC.

\paragraph{Broader Impact}

LLM-based rewriting improves code retrieval but inherits the
rewriter's biases and hallucination risk: a paraphrase that silently
changes semantics can mislead downstream retrieval and any consuming
system (e.g., code completion, security audit, or program repair).
Our offline (C) pipeline partially mitigates this by allowing human
review of the rewritten corpus before deployment. We recommend that
practitioners (a) audit a random sample of rewrites for semantic
drift, (b) retain pointers from rewritten entries to the original
source, and (c) prefer QC-NL only when retrieval gains outweigh the
compute cost and hallucination risk for the target application.

\bibliography{bibl}
\bibliographystyle{bibl}

%%%%%%%%%%%%%%%%%%%%%%%%%%%%%%%%%%%%%%%%%%%%%%%%%%%%%%%%%%%%

\appendix
\newpage
\section{Technical Appendices and Supplementary Material}
\label{appendix}

\begin{table*}[!ht]
    \caption{\textbf{Per-benchmark NDCG@10 for all (encoder,
      technique, augmentation) combinations} on the four code-to-code
      and text-to-code CoIR tasks.  \textit{Baseline} denotes the
      unmodified corpus and query (shown in grey).  \textit{QC} =
      query+corpus rewriting; \textit{C} = corpus-only rewriting.
      Bold marks the best configuration per encoder per column.}

    \resizebox{\textwidth}{!}{\centering
    \begin{tabular}{ccccccc|c}

        \midrule
        Model & Augmented &Technique & Apps & CosQA & CT-Contest & CT-DL & Avg. \\

        \midrule
        \textsc{CodeXembed} &QC& Rephrasing &\textbf{0.62} & 0.20 &0.76  &0.28 &0.46  \\
        \textsc{CodeXembed}  &QC&PseudoCode& 0.56& 0.25&0.82 & 0.31 & 0.48\\
        \textsc{CodeXembed}  &QC&NL& 0.54& 0.22&\textbf{0.87} & \textbf{0.32} & \textbf{0.49}\\
        \textsc{CodeXembed} &C&Rephrasing & 0.56 & 0.31&0.69 &0.28  & 0.46\\
        \textsc{CodeXembed}  &C&PseudoCode& 0.55& 0.28& 0.68&0.27 & 0.44 \\
        \textsc{CodeXembed}  &C&NL& 0.49& 0.25& 0.66&  0.23& 0.41\\ \rowcolor{gray!12}
        \textsc{CodeXembed} &X&Baseline& 0.48 & \textbf{0.34} & 0.70 & 0.33 &0.46  \\
        \midrule
        \textsc{MoSE-18} &QC&Rephrasing & 0.16  & \textbf{0.15} & 0.44  & 0.29 & 0.26 \\
        \textsc{MoSE-18}  &QC& PseudoCode&\textbf{0.18}&\textbf{0.15} &0.54 & 0.29 & 0.29 \\
        \textsc{MoSE-18}  &QC& NL&0.13&0.10 &\textbf{0.74} & \textbf{0.36} &  \textbf{0.33}\\
        \textsc{MoSE-18} &C&Rephrasing &  0.04 & 0.04 & 0.30  &  0.24 & 0.15 \\
        \textsc{MoSE-18}  &C&PseudoCode&0.05 &0.03 &0.31 &0.20  & 0.15 \\
        \textsc{MoSE-18}  &C& NL&0.04& 0.03& 0.20&0.07  & 0.08 \\ \rowcolor{gray!12}
        \textsc{MoSE-18} &X&Baseline&  0.01 & 0.04 & 0.23  & 0.20   & 0.12  \\        
                \midrule
        \textsc{Qwen3-Embedding-0.6B}  &QC&Rephrasing & 0.69  & 0.26&0.87& 0.31 & 0.53\\
        \textsc{Qwen3-Embedding-0.6B}  &QC&PseudoCode& 0.58&0.29 &0.88 & 0.35 &0.52 \\
        \textsc{Qwen3-Embedding-0.6B}  &QC&NL&0.56 &0.28 &\textbf{ 0.90}& \textbf{0.34} & 0.52\\
        \textsc{Qwen3-Embedding-0.6B}  &C&Rephrasing & 0.64  & 0.32& 0.86& 0.29 &0.53\\
        \textsc{Qwen3-Embedding-0.6B}   &C&PseudoCode&0.53 &0.31 & 0.83&0.31  & 0.49 \\
        \textsc{Qwen3-Embedding-0.6B}  &C&NL& 0.40& 0.30&0.86 & 0.28 & 0.46\\ \rowcolor{gray!12}
        \textsc{Qwen3-Embedding-0.6B}  &X&Baseline&  \textbf{0.70} & \textbf{0.38}& 0.84 & 0.31 &\textbf{0.56} \\
                \midrule
        \textsc{Unixcoder base} &QC&Rephrasing &\textbf{0.28} & 0.17&0.57 & 0.31 & \textbf{0.33} \\
        \textsc{Unixcoder base}  &QC&PseudoCode& 0.21&0.15 & 0.62&\textbf{0.31} & 0.32\\
        \textsc{Unixcoder base}  &QC&NL& 0.19&0.16 &\textbf{0.67} & 0.30& \textbf{0.33} \\
        \textsc{Unixcoder base} &C&Rephrasing & 0.04 & 0.20&0.42 & 0.28 &0.24\\
        \textsc{Unixcoder base}  &C&PseudoCode&0.09 &0.18 &0.44& 0.25& 0.24\\
        \textsc{Unixcoder base}  &C&NL&0.03 &0.15 &0.27 & 0.15& 0.15\\ \rowcolor{gray!12}
        \textsc{Unixcoder base} &X&Baseline&0.02 & \textbf{0.24}&0.45 & 0.31& 0.25 \\
                \midrule
        \textsc{e5-base-v2} &QC&Rephrasing & \textbf{0.35} &  0.19& 0.69 & 0.22 & 0.36 \\
        \textsc{e5-base-v2}  &QC&PseudoCode& 0.31 &0.17 & 0.73& 0.30 & 0.38\\
        \textsc{e5-base-v2}  &QC&NL& 0.35 &0.20 & \textbf{0.85}& \textbf{0.31} & \textbf{0.43}\\
        \textsc{e5-base-v2} &C&Rephrasing & 0.19 & 0.27 & 0.63 & 0.16 & 0.31\\
        \textsc{e5-base-v2}  &C&PseudoCode& 0.21 & \textbf{0.28}& 0.60& 0.16&0.31 \\
        \textsc{e5-base-v2}  &C&NL& 0.20 & 0.21&0.60 & 0.12 &0.28 \\ \rowcolor{gray!12}
        \textsc{e5-base-v2} &X&Baseline& 0.11 & 0.31 & 0.61 & 0.21 &0.31 \\
        \bottomrule

    \end{tabular}}

    \label{table:resultst2c}
\end{table*}

\begin{table*}[!ht]
    \caption{\textbf{Per-benchmark NDCG@10 on the two hybrid CoIR
      tasks} (StackOverflow-QA, CodeFeedback-MT), where queries and
      documents natively mix natural language and code.  Format as in
      Table~\ref{table:resultst2c}.  \textit{Baseline} rows are shown
      in grey.}

    \resizebox{\textwidth}{!}{\centering
    \begin{tabular}{cccccc}

        \midrule
        Model & Augmented & Technique & StackOverflow-QA & CodeFeedback-MT \\
        \midrule
         \textsc{CodeXembed}& QC&Rephrasing &0.46 &0.11  \\
        \textsc{CodeXembed}& QC&PseudoCode & 0.47&0.11  \\
        \textsc{CodeXembed}&QC &NL &0.47 & 0.11 \\  
        \textsc{CodeXembed}& C&Rephrasing & 0.47& 0.09 \\
        
        \textsc{CodeXembed}& C&PseudoCode & 0.47& 0.09 \\
        \textsc{CodeXembed}&C &NL & 0.46&0.08  \\
        \rowcolor{gray!12} \textsc{CodeXembed}&X &Baseline & 0.49&0.08  \\
        \midrule
         \textsc{MoSE-18}& QC&Rephrasing & 0.29&0.09  \\
        \textsc{MoSE-18}& QC&PseudoCode &0.36 &0.09  \\
        \textsc{MoSE-18}&QC &NL &0.32&0.09  \\  
        \textsc{MoSE-18}& C&Rephrasing & 0.24&0.06  \\
        
        \textsc{MoSE-18}& C&PseudoCode & 0.25&0.06  \\
        \textsc{MoSE-18}&C &NL &0.18 &0.05  \\
        \rowcolor{gray!12} \textsc{MoSE-18}&X &Baseline &0.24 & 0.02 \\
        \midrule
         \textsc{Qwen3-Embedding-0.6B}& QC&Rephrasing & 0.48&0.13  \\
        \textsc{Qwen3-Embedding-0.6B}& QC&PseudoCode &0.48 &0.12  \\
        \textsc{Qwen3-Embedding-0.6B}&QC &NL &0.48&0.12  \\  
        \textsc{Qwen3-Embedding-0.6B}& C&Rephrasing &0.48 &0.10  \\
        
        \textsc{Qwen3-Embedding-0.6B}& C&PseudoCode & 0.48& 0.10 \\
        \textsc{Qwen3-Embedding-0.6B}&C &NL &0.48 &0.10  \\
        \rowcolor{gray!12} \textsc{Qwen3-Embedding-0.6B}&X &Baseline & 0.49& 0.11 \\
        \midrule
         \textsc{UniXcoder Base}& QC&Rephrasing &0.33 &0.08  \\
        \textsc{UniXcoder Base}& QC&PseudoCode &0.33 &0.08  \\
        \textsc{UniXcoder Base}&QC &NL &0.36&0.07  \\  
        \textsc{UniXcoder Base}& C&Rephrasing &0.28 & 0.05 \\
        
        \textsc{UniXcoder Base}& C&PseudoCode & 0.29&0.06  \\
        \textsc{UniXcoder Base}&C &NL &0.17 &0.06  \\
        \rowcolor{gray!12} \textsc{UniXcoder Base}&X &Baseline &0.33 &0.06  \\
        \midrule
         \textsc{E5-Base-V2}& QC&Rephrasing &0.44 &0.11  \\
        \textsc{E5-Base-V2}& QC&PseudoCode & 0.45&0.11  \\
        \textsc{E5-Base-V2}&QC &NL &0.45&0.11  \\  
        \textsc{E5-Base-V2}& C&Rephrasing &0.45 &0.08  \\
        
        \textsc{E5-Base-V2}& C&PseudoCode &0.45 &0.11  \\
        \textsc{E5-Base-V2}&C &NL &0.43 &0.07  \\
        \rowcolor{gray!12} \textsc{E5-Base-V2}&X &Baseline &0.48 & 0.07 \\
        \bottomrule
        
    \end{tabular}}

    \label{table:resultshyb}
\end{table*}

\begin{table}[!ht]
\centering
\caption{%
  Tokenisation statistics averaged across tasks for each encoder--strategy
  combination ($n{=}4$ tasks per cell).
  \textit{Vocab} is the encoder vocabulary size;
  \textit{Unique} is the number of distinct tokens observed in the corpus;
  $H$ is the token unigram entropy (bits);
  \textit{TTR} is the type--token ratio;
  \textit{Top-20\%} is the fraction of total token mass carried by the most
  frequent 20\% of types (a measure of distributional skew);
  \textit{Hapax\%} is the proportion of token types appearing exactly once
  (a measure of lexical richness at the tail).
  Higher $H$ and \textit{Hapax\%} indicate a flatter, richer distribution;
  higher \textit{Top-20\%} indicates greater concentration on frequent types.
  $\Delta H$ is computed relative to the per-encoder baseline row
  (shown in grey).
  NL rewriting consistently produces the largest entropy gain and hapax rate
  across all encoders, while pseudo-code maximises lexical breadth
  (unique types) without a proportional increase in tail richness;
  the Top-20\% mass falls by up to 21 pp relative to baseline,
  confirming a systematic redistribution toward the long tail
  regardless of encoder vocabulary size.
}
\label{tab:tokenisation}
\small
\setlength{\tabcolsep}{5pt}
\renewcommand{\arraystretch}{1.15}
\begin{tabular}{llrrrrrrr}
\toprule
\textbf{Encoder}
  & \textbf{Strategy}
  & \textbf{Vocab}
  & \textbf{Unique}
  & $H$ \textbf{(bits)}
  & $\Delta H$
  & \textbf{TTR}
  & \textbf{Top-20\%}
  & \textbf{Hapax\%} \\
\midrule
\rowcolor{gray!12}
CodeXEmbed  & Baseline          & 30{,}522  & 1{,}078  & 7.14 & ---         & 0.054 & 52.6\% & 27.8\% \\
CodeXEmbed  & NL                & 30{,}522  & 1{,}624  & 8.52 & $+1.38$     & 0.138 & 35.7\% & 39.9\% \\
CodeXEmbed  & Pseudo            & 30{,}522  & 2{,}128  & 8.06 & $+0.92$     & 0.049 & 41.4\% & 27.9\% \\
CodeXEmbed  & Rephrase          & 30{,}522  & 1{,}718  & 7.74 & $+0.60$     & 0.048 & 44.8\% & 31.1\% \\
\midrule
\rowcolor{gray!12}
E5-base-v2  & Baseline          & 30{,}522  & 1{,}078  & 7.14 & ---         & 0.054 & 52.6\% & 27.8\% \\
E5-base-v2  & NL                & 30{,}522  & 1{,}624  & 8.52 & $+1.38$     & 0.138 & 35.7\% & 39.9\% \\
E5-base-v2  & Pseudo            & 30{,}522  & 2{,}128  & 8.06 & $+0.92$     & 0.049 & 41.4\% & 27.9\% \\
E5-base-v2  & Rephrase          & 30{,}522  & 1{,}718  & 7.74 & $+0.60$     & 0.048 & 44.8\% & 31.1\% \\
\midrule
\rowcolor{gray!12}
MoSE-18     & Baseline          & 49{,}152  & 1{,}636  & 8.03 & ---         & 0.088 & 41.5\% & 33.3\% \\
MoSE-18     & NL                & 49{,}152  & 1{,}940  & 8.89 & $+0.86$     & 0.169 & 31.0\% & 45.4\% \\
MoSE-18     & Pseudo            & 49{,}152  & 3{,}447  & 8.76 & $+0.73$     & 0.077 & 37.7\% & 34.3\% \\
MoSE-18     & Rephrase          & 49{,}152  & 2{,}622  & 8.45 & $+0.42$     & 0.072 & 37.8\% & 35.2\% \\
\midrule
\rowcolor{gray!12}
Qwen3-Emb   & Baseline          & 151{,}643 & 2{,}031  & 8.59 & ---         & 0.117 & 34.8\% & 36.6\% \\
Qwen3-Emb   & NL                & 151{,}643 & 1{,}955  & 8.88 & $+0.29$     & 0.176 & 31.5\% & 47.5\% \\
Qwen3-Emb   & Pseudo            & 151{,}643 & 3{,}874  & 9.08 & $+0.49$     & 0.088 & 33.6\% & 36.8\% \\
Qwen3-Emb   & Rephrase          & 151{,}643 & 3{,}044  & 8.97 & $+0.38$     & 0.088 & 31.4\% & 35.7\% \\
\midrule
\rowcolor{gray!12}
UniXcoder   & Baseline          & 51{,}416  & 1{,}701  & 7.86 & ---         & 0.086 & 44.2\% & 33.8\% \\
UniXcoder   & NL                & 51{,}416  & 1{,}968  & 8.94 & $+1.08$     & 0.171 & 30.8\% & 45.0\% \\
UniXcoder   & Pseudo            & 51{,}416  & 3{,}512  & 8.80 & $+0.94$     & 0.079 & 36.7\% & 34.2\% \\
UniXcoder   & Rephrase          & 51{,}416  & 2{,}670  & 8.38 & $+0.52$     & 0.074 & 38.7\% & 35.0\% \\
\bottomrule
\end{tabular}
\end{table}

\begin{table*}[!ht]
    \caption{\textbf{Per-encoder in-vitro results on
      codetrans-contest} (NDCG@10) for four rewriter sizes from the
      \textsc{Qwen2.5-Coder-Instruct} family (1.5B--14B).  Larger
      rewriters consistently yield higher retrieval quality across
      all three strategies and all five encoders, confirming that
      rewriter capacity is a primary driver of downstream gains.}

    \resizebox{\textwidth}{!}{\centering
    \begin{tabular}{cccccc}

        \midrule
        Model  &Technique & 1.5B & 3B & 7B & 14B  \\

        \midrule
        \textsc{CodeXembed} & Rephrasing & 0.63& 0.58 & 0.71 & 0.71  \\
        \textsc{CodeXembed}  &PseudoCode& 0.55 &0.45 &0.65 & 0.67 \\
        \textsc{CodeXembed}  &NL& 0.74&0.82 &0.80 & 0.82 \\

        \midrule
        \textsc{MoSE-18} &Rephrasing & 0.36  & 0.39& 0.37  & 0.38  \\
        \textsc{MoSE-18} &PseudoCode &  0.35 &0.38 &  0.44 & 0.33  \\
        \textsc{MoSE-18} &NL &  0.44 & 0.51& 0.58  & 0.58  \\
        \midrule
        \textsc{Qwen3-Emb} &Rephrasing &   0.81&0.83 & 0.85  & 0.84  \\
        \textsc{Qwen3-Emb} &PseudoCode & 0.83  & 0.81& 0.84  & 0.82  \\
        \textsc{Qwen3-Emb} &NL & 0.81  &0.84 &  0.85 &  0.83 \\
        
        \midrule
        \textsc{UniXcoder} &Rephrasing &  0.52 & 0.56&  0.56 &  0.55 \\
        \textsc{UniXcoder} &PseudoCode &  0.48 &0.53 & 0.54  & 0.55  \\
        \textsc{UniXcoder} &NL & 0.43  &0.53 &  0.53 &  0.58 \\
        \midrule
        \textsc{E5-Base-V2} &Rephrasing & 0.49   & 0.53&  0.59 &  0.68 \\
        \textsc{E5-Base-V2} &PseudoCode & 0.36  &0.34 & 0.50  &  0.58 \\
        \textsc{E5-Base-V2} &NL &  0.59 & 0.72&  0.73 & 0.74 \\
        \bottomrule

    \end{tabular}}

    \label{table:in_vitro}
\end{table*}

\begin{table*}[!ht]
    \caption{\textbf{Cross-rewriter per-encoder NDCG@10 on codetrans-contest.}
      Comparison across three rewriters spanning three independent model
      families: Qwen3-Coder-30B (Qwen), DeepSeek-Coder-V2-Lite-Instruct
      (DeepSeek), and Codestral-22B (Mistral).  All experiments use
      \textsc{QC-manipulation}.  NL rewriting is the best or competitive
      strategy for every encoder under at least two of three rewriters,
      confirming the cross-family robustness of our headline finding.}
    \resizebox{\textwidth}{!}{\centering
    \begin{tabular}{ccccc}
        \toprule
        Model  & Technique & Qwen3-Coder-30B & DeepSeek-V2-Lite & Codestral-22B  \\
        \midrule
        \textsc{CodeXembed} & Rephrasing & 0.76 & 0.69 & \textbf{0.83} \\
        \textsc{CodeXembed} & PseudoCode & 0.82 & 0.57 & 0.78 \\
        \textsc{CodeXembed} & NL         & \textbf{0.87} & \textbf{0.77} & 0.80 \\
        \rowcolor{gray!12}
        \textsc{CodeXembed} & Baseline   & 0.70 & -- & -- \\
        \midrule
        \textsc{MoSE-18} & Rephrasing & 0.44 & 0.37 & \textbf{0.58} \\
        \textsc{MoSE-18} & PseudoCode & 0.54 & 0.39 & 0.46 \\
        \textsc{MoSE-18} & NL         & \textbf{0.74} & \textbf{0.51} & 0.56 \\
        \rowcolor{gray!12}
        \textsc{MoSE-18} & Baseline   & 0.23 & -- & -- \\
        \midrule
        \textsc{Qwen3-Emb} & Rephrasing & 0.87 & 0.79 & 0.86 \\
        \textsc{Qwen3-Emb} & PseudoCode & 0.88 & \textbf{0.81} & 0.85 \\
        \textsc{Qwen3-Emb} & NL         & \textbf{0.90} & \textbf{0.81} & \textbf{0.87} \\
        \rowcolor{gray!12}
        \textsc{Qwen3-Emb} & Baseline   & 0.84 & -- & -- \\
        \midrule
        \textsc{UniXcoder} & Rephrasing & 0.57 & 0.48 & \textbf{0.68} \\
        \textsc{UniXcoder} & PseudoCode & 0.62 & 0.46 & 0.57 \\
        \textsc{UniXcoder} & NL         & \textbf{0.67} & \textbf{0.53} & 0.61 \\
        \rowcolor{gray!12}
        \textsc{UniXcoder} & Baseline   & 0.45 & -- & -- \\
        \midrule
        \textsc{E5-Base-V2} & Rephrasing & 0.69 & 0.58 & \textbf{0.75} \\
        \textsc{E5-Base-V2} & PseudoCode & 0.73 & 0.41 & 0.62 \\
        \textsc{E5-Base-V2} & NL         & \textbf{0.85} & \textbf{0.63} & \textbf{0.75} \\
        \rowcolor{gray!12}
        \textsc{E5-Base-V2} & Baseline   & 0.61 & -- & -- \\
        \bottomrule
    \end{tabular}}
    \label{table:cross_contest}
\end{table*}

\begin{table*}[!ht]
    \caption{\textbf{Cross-rewriter per-encoder NDCG@10 on cosqa.}
      On this NL-heavy text-to-code benchmark, no rewriting strategy
      improves over the unmodified baseline under any rewriter,
      confirming that the failure of rewriting on NL-dominant queries
      is an intrinsic property of the task rather than an artifact of
      the Qwen rewriter.  This negative result is correctly anticipated
      by our $\Delta H$ diagnostic (\S\ref{subsec:cross-rewriter}),
      which is near-zero or negative on cosqa for all three rewriters.}
    \resizebox{\textwidth}{!}{\centering
    \begin{tabular}{ccccc}
        \toprule
        Model  & Technique & Qwen3-Coder-30B & DeepSeek-V2-Lite & Codestral-22B \\
        \midrule
        \textsc{CodeXembed} & Rephrasing & 0.20 & 0.19 & 0.20 \\
        \textsc{CodeXembed} & PseudoCode & 0.25 & 0.16 & 0.21 \\
        \textsc{CodeXembed} & NL         & 0.22 & 0.18 & 0.22 \\
        \rowcolor{gray!12}
        \textsc{CodeXembed} & Baseline   & \textbf{0.34} & -- & -- \\
        \midrule
        \textsc{MoSE-18} & Rephrasing & 0.15 & 0.13 & 0.12 \\
        \textsc{MoSE-18} & PseudoCode & 0.15 & 0.10 & 0.13 \\
        \textsc{MoSE-18} & NL         & 0.10 & 0.12 & 0.09 \\
        \rowcolor{gray!12}
        \textsc{MoSE-18} & Baseline   & 0.04 & -- & -- \\
        \midrule
        \textsc{Qwen3-Emb} & Rephrasing & 0.26 & 0.20 & 0.23 \\
        \textsc{Qwen3-Emb} & PseudoCode & 0.29 & 0.17 & 0.22 \\
        \textsc{Qwen3-Emb} & NL         & 0.28 & 0.21 & 0.22 \\
        \rowcolor{gray!12}
        \textsc{Qwen3-Emb} & Baseline   & \textbf{0.38} & -- & -- \\
        \midrule
        \textsc{UniXcoder} & Rephrasing & 0.17 & 0.16 & 0.14 \\
        \textsc{UniXcoder} & PseudoCode & 0.15 & 0.11 & 0.16 \\
        \textsc{UniXcoder} & NL         & 0.16 & 0.12 & 0.15 \\
        \rowcolor{gray!12}
        \textsc{UniXcoder} & Baseline   & \textbf{0.24} & -- & -- \\
        \midrule
        \textsc{E5-Base-V2} & Rephrasing & 0.19 & 0.15 & 0.12 \\
        \textsc{E5-Base-V2} & PseudoCode & 0.17 & 0.12 & 0.15 \\
        \textsc{E5-Base-V2} & NL         & 0.20 & 0.14 & 0.19 \\
        \rowcolor{gray!12}
        \textsc{E5-Base-V2} & Baseline   & \textbf{0.31} & -- & -- \\
        \bottomrule
    \end{tabular}}
    \label{table:cross_cosqa}
\end{table*}

%%%%%%%%%%%%%%%%%%%%%%%%%%%%%%%%%%%%%%%%%%%%%%%%%%%%%%%%%%%%

% \clearpage
% \newpage
% \input{checklist.tex}

\end{document}